\documentclass[epj,nopacs]{svjour}

\usepackage{graphics}

\begin{document}

\title{Evidence for Factorization in Three-body
$\overline B\to D^{(*)}\,K^-\,K^0$ Decays}


\author{Chun-Khiang Chua\inst{1} \and George W. S. Hou\inst{2} \and Shiue-Yuan Shiau\inst{3}
\and Shang-Yuu Tsai\inst{2}
}                     


\institute{Institute of Physics, Academia Sinica,\\
Taipei, Taiwan 115, Republic of China \and Department of Physics,
National Taiwan University,\\
Taipei, Taiwan 10764, Republic of China \and Department of
Physics, University of Wisconsin,\\ Madison, WI 53706, USA}
\date{\today} 

\abstract{Motivated by experimental results on $\overline B\to
D^{(*)}K^-K^{0}$, we use a factorization approach to study these
decays. Two mechanisms concerning kaon pair production arise:
current-produced (from vacuum) and transition (from the $B$
meson). The kaon pair in the $\overline B {}^0\to D^{(*)+}K^-K^0$
decays can be produced only by the vector
current~(current-produced), whose matrix element can be extracted
from $e^+e^-\to K\overline K$ processes via isospin relations. The
decay rates obtained this way are in good agreement with
experiment. The $B^-\to D^{(*)0}K^- K^0$ decays involve both
current-produced and transition processes. By using QCD counting
rules and the measured $B^-\to D^{(*)0} K^- K^0$ decay rates, the
measured decay spectra can be understood.
} 

\maketitle

\section{Introduction}
\label{intro}

The $\overline B\to D^{(*)}K^-K^{(*)0}$ decays have been observed
for the first time by the Belle
Collaboration~\cite{Drutskoy:2002ib}, with branching fractions at
the level of $10^{-4}-10^{-3}$. Angular analysis reveals that
$K^-K^0$ and $K^-K^{*0}$ are dominantly $J^P=1^-$ and $1^+$,
respectively. While there is no sign of decay via resonance for
the $K^-K^0$ pair, data suggest a dominant $a_1(1260)$ resonance
contribution in the production of the $K^-K^{*0}$. The
$K^-K^{(*)0}$ mass spectra are peaking near threshold.

The near-threshold peaking of the $K^-K^{(*)0}$ mass spectra
suggests a {\it quasi two-body} process where the colinear
$K^-K^{(*)0}$ recoil against the $D^{(*)}$ meson. This is
suggestive to apply fatorization to the three-body
case~\cite{Chua:2002pi}. Two kinds of decay amplitudes arise due
to the flavor structures of the $D^{(*)}$ mesons:
$D^{(*)+}K^-K^{(*)0}$ involves $\langle K^-K^{(*)0}|V-A|0\rangle$,
with $K^-K^{(*)0}$ produced by a weak $V-A$ current;
$D^{(*)0}K^-K^{(*)0}$ involves $\langle
K^-K^{(*)0}|V-A|B^-\rangle$, where $B^-$ goes into $K^-K^{(*)0}$
via a weak current.

In $\langle K^-K^0|V-A|0\rangle$, the $K^-K^0$ can only be
produced by the vector current, and should be dominantly $1^-$. By
isospin rotation, the kaon weak form factor $\langle
K^-K^0|V|0\rangle$ can therefore be related to the kaon
electromagnetic~(EM) form factors in $e^+e^-$ annihilation, where
much data exist. One can then calculate the rate without any
tuning parameters. The predicted $K^-K^0$ mass spectrum can be
shown to have a peak near threshold, which arises from the kaon
form factor and can be checked by experiment.

In $B^-\to D^{(*)0} K^- K^0$ decays, the $K^-K^0$ can also be
produced by a current that induces $B^-\to K^-K^0$ transitions.
The relevant matrix element $\langle K^-K^0|V-A|B^-\rangle$ is
parameterized by several {\it unknown} form factors, due to which
the rate cannot be calculated. Nevertheless, by using a naive
parametrization based on QCD counting rules~\cite{Brodsky:1974vy},
the parameters in these unknown form factors can be determined
from the decay rates, and the decay spectra can be obtained, which
again have threshold enhancement as closely related to QCD
counting rules, and can be tested experimentally.

The $K^-K^{*0}$ in $\overline B {}^0\to D^{(*)+}K^-K^{*0}$ can
only be produced by a weak current. The experimental observation
that $K^-K^{*0}$ is in $1^+$ suggests a dominant {\it axial}
current contribution. Although the decay rate cannot be calculated
due to the absence of data of the $K^-K^{*0}$ axial form factors,
it has been proposed that one can extract the $K^-K^{*0}$ axial
form factors given the $K^-K^{*0}$ mass spectrum~\cite{DKKstar}.

In what follows, we shall concentrate on decays that involve only
$K^-K^0$. In the next section we will introduce the relevant
formalism and describe the numerical results. A discussion will be
given in the last section where the conclusion is drawn.
%
\begin{figure}
\resizebox{0.43\textwidth}{!}{
\includegraphics{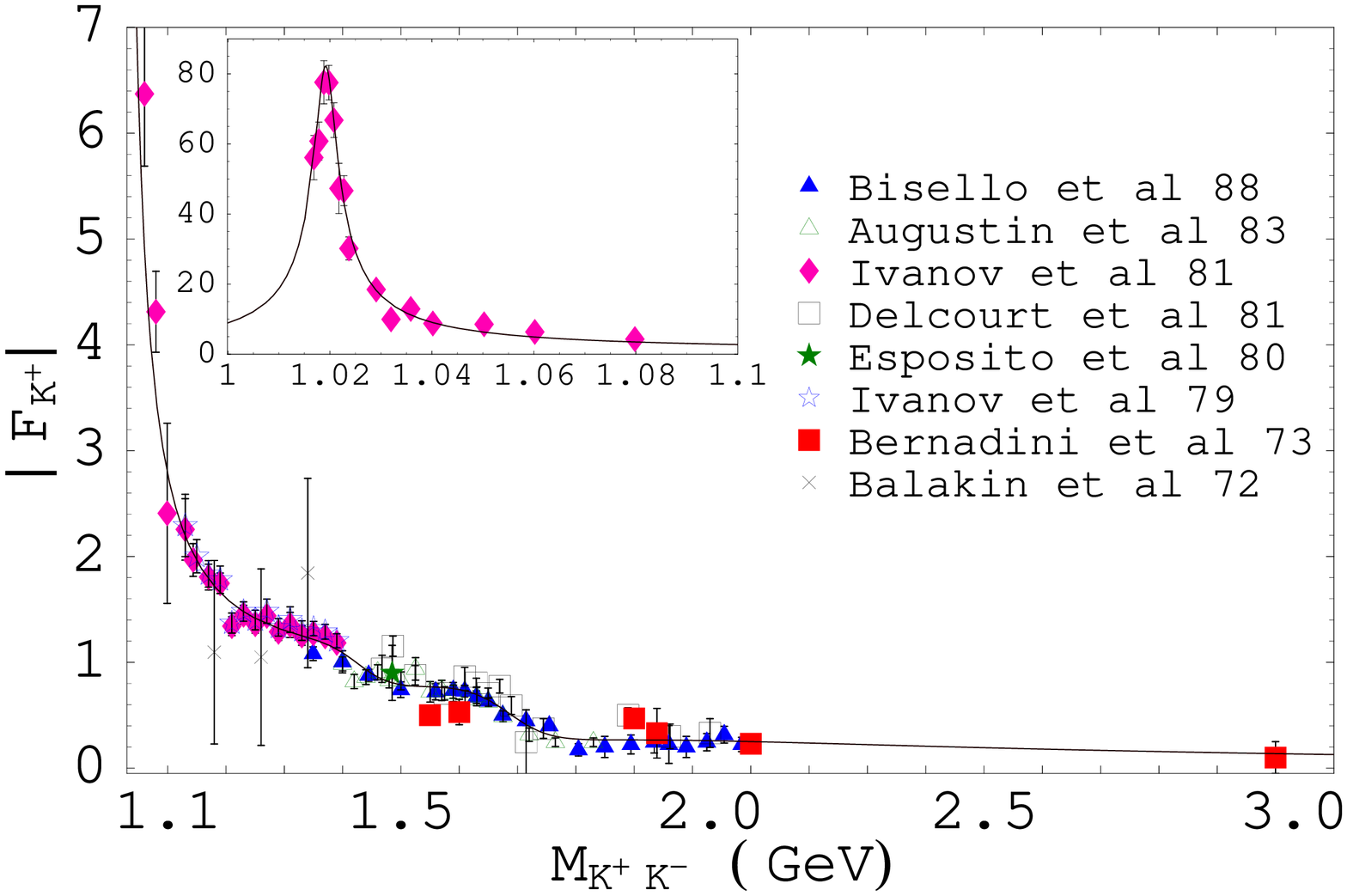} 
} \resizebox{0.43\textwidth}{!}{
\includegraphics{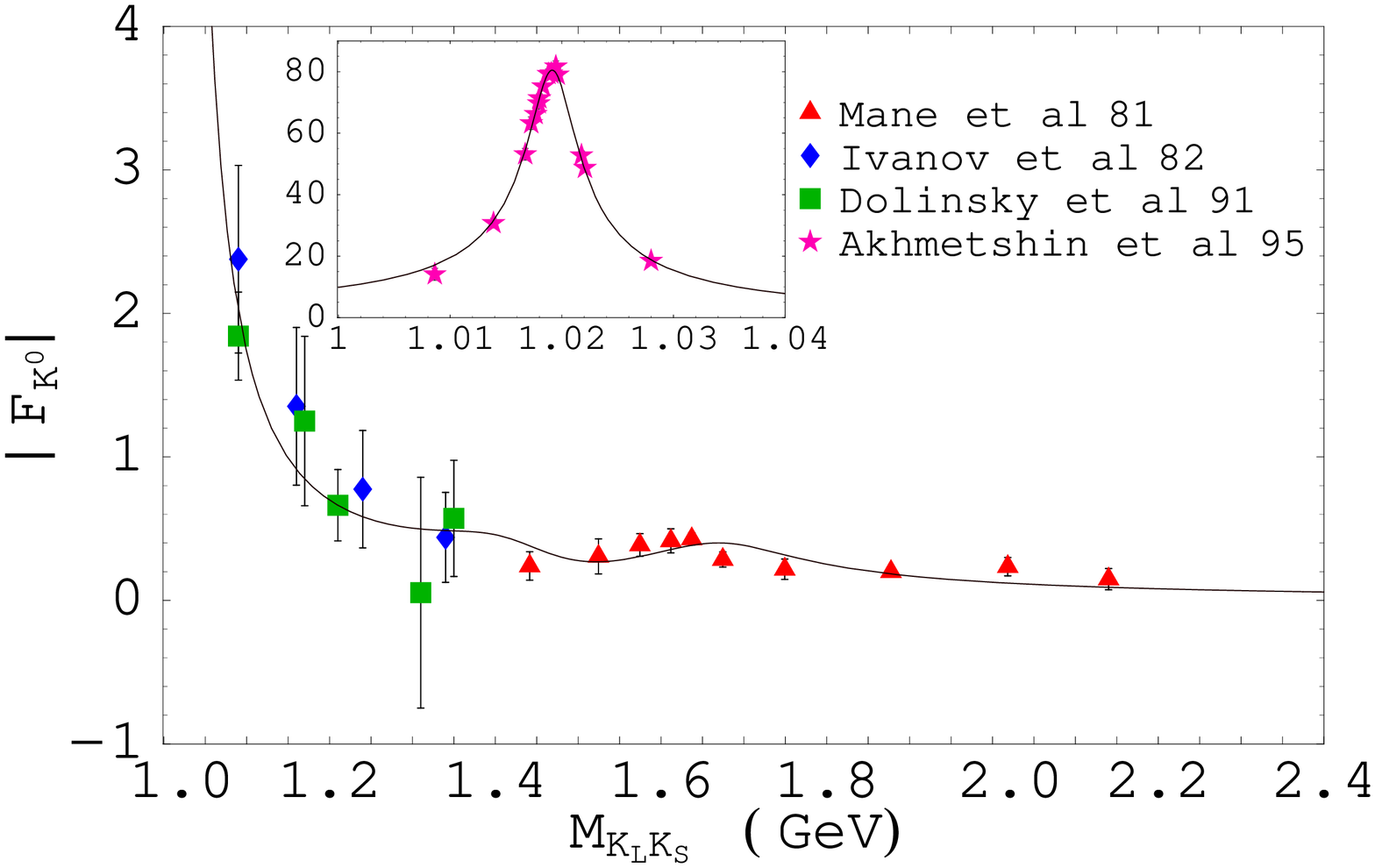}
}
\caption{\small {
 Fit to timelike $|F_{K^+}|$ (upper) and $|F_{K^0}|$ (lower) form factor data,
 where the inset is for the $\phi$ region.
 }}\label{fig:KKff}
\end{figure}
%

\section{Factorization Formalism}
\label{sec:formalism}
Starting with the relevant effective Hamiltonian, and the
factorization ansatz, one arrives at~\cite{Chua:2002pi}
\begin{eqnarray}
{\mathcal A}(D^{(*)+}K^-K^{(*)0})&=& \frac{G_F}{\sqrt2} V_{cb}
V_{ud}^*
\,a_1 \langle D^{(*)+}|(\bar c b)_{V-A}|\overline B {}^0\rangle \nonumber\\
&& \times \langle K^-K^{(*)0}|(\bar d u)_V|0\rangle,
\label{eq:current-produced}\\
{\mathcal A}(D^{(*)0}K^-K^{(*)0})&=& \frac{G_F}{\sqrt2} V_{cb}
V_{ud}^* \big[
 a_1 \langle D^{(*)0}|(\bar c b)_{V-A}|B^-\rangle \nonumber\\
&& \times \langle K^-K^{(*)0}|(\bar d u)_V|0\rangle
\nonumber\\
&& +a_2 \langle K^-K^{(*)0}|(\bar d b)_{V-A}|B^-\rangle \nonumber\\
&& \times \langle D^{(*)0}|(\bar c u)_{V-A}|0\rangle \big],
\label{eq:transition}
\end{eqnarray}
in which
\begin{equation}
\langle K^-(p_{K^-})K^0(p_{K^0})\left|V^\mu\right|0\rangle=
(p_{K^-}-p_{K^0})^\mu F_1^{KK}(q^2) \label{eq:Jff}
\end{equation}
in the isospin limit, and
\begin{eqnarray}
&&\langle K^-(p_{K^-})K^0(p_{K^0})|(V-A)_\mu|
B^-(p_B)\rangle\nonumber\\
 &&\qquad= i w_-(q^2) (p_{K^-}-p_{K^0})_\mu  \nonumber\\  
&&\qquad\quad+\,h(q^2)\,\epsilon_{\mu\nu\alpha\beta}\,p_B^\nu
q^\alpha (p_{K^-}-p_{K^0})^\beta, \label{eq:Tff}
\end{eqnarray}
where $q\equiv p_{K^-}+p_{K^0}$. The fact that $K^-K^0$ is in
$1^-$ has been taken into account in the above parametrizations.
The $\langle D^{(*)}|(\bar c b)_{V-A}|\overline B \rangle$ is the
same as in two-body cases and we adopt both the
BSW~\cite{BSW:physC29} and the MS~\cite{Melikhov:2000yu} models
for comparison.

\begin{figure}
\centerline{ \resizebox{0.35\textwidth}{!}{
\includegraphics{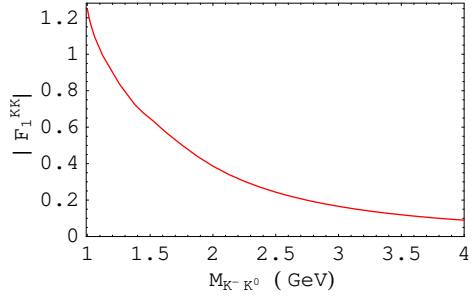}}}
\caption{The kaon weak vector form factor
$F^{KK}_1$.}\label{fig:KweakFF}
\end{figure}

The kaon weak vector form factor $F^{KK}_1$ is related to its EM
partners via the isospin relation
\begin{equation}
F^{KK}_1(q^2)=F_{K^+}(q^2)-F_{K^0}(q^2), \label{eq:iso_relation}
\end{equation}
where $F_{K^+}$, $F_{K^0}$ are the EM form factors of the charged
and neutral kaons, respectively. By fitting to the EM data, one
can obtain the kaon EM form factors and hence the weak vector form
factor~\cite{Chua:2002pi}, as shown in Figs.~\ref{fig:KKff} and
\ref{fig:KweakFF}. Readers are referred to \cite{Chua:2002pi} for
detailed discussion on the fitting of kaon EM form factors.
%
%

For the unknown form factors in Eq.~\ref{eq:Tff}, we take the
following naive parametrization~\cite{Chua:2002pi}
\begin{equation}
w_-(t)=\frac{c_{w}}{t^2},\qquad h(t)=\frac{c_{h}}{t^2},
\label{eq:transitionFF}
\end{equation}
where $c_{w,h}$ are free parameters to be fitted by data. The
$1/t^2$ arises from the minimum number of hard gluons to produce
an energetic kaon pair from a decaying $B$ meson, which
characterizes the asymptotic behavior of the form factor.

\begin{table}
\caption{${\cal B}(\overline B^0\to D^{(*)+}K^-K^0)$
 in units of $10^{-4}$.
 }\label{tab:BRD+KK}
\begin{tabular}{lccc}
\hline\noalign{\smallskip}
 & MS &  BSW &  Experiment~\cite{Drutskoy:2002ib}  \\
\noalign{\smallskip}\hline\noalign{\smallskip} $ \overline
B{}^0\to D^+K^-K^0$  & $1.67^{+0.24}_{-0.21}$
&$1.54^{+0.22}_{-0.20}$ & $1.6\pm0.8\pm0.3$ \\

  & & & $<3.1$~($90\%$~CL)\\
$\overline B {}^0\to D^{*+}K^-K^0$
&$2.8^{+0.30}_{-0.36}$&$3.05^{+0.32}_{-0.39}$& $2.0\pm1.5\pm0.4$
\\
 & & & $<4.7$~($90\%$~CL) \\
\noalign{\smallskip}\hline
\end{tabular}
\end{table}

In Table~\ref{tab:BRD+KK} we show the calculated branching
fractions of the $\overline B {}^0\to D^{(*)+}K^-K^0$ modes. One
can see that the results are in very good agreement with
experiment. In the absence of any tuning parameters in our
formalism for these decay modes, the agreement between experiment
and the model actually provides {\it evidence} for factorization!

\begin{figure}
\centerline{\resizebox{0.4\textwidth}{!}{%
  \includegraphics{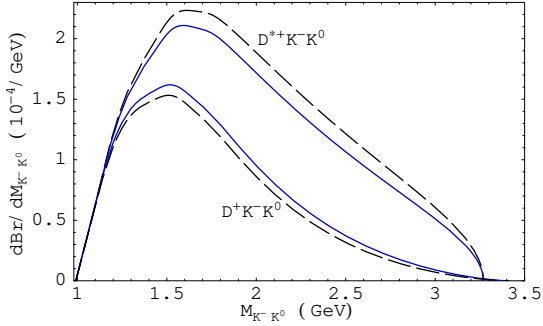}
}} \caption{ The $K^-K^0$ mass spectrum for $\overline B {}^0\to
D^+K^-K^0$ (lower) and $D^{*+}K^-K^0$ (upper), where
solid~(dashed) line stands for using the MS~(BSW) hadronic form
factors.}\label{fig:D+KK}
\end{figure}

On the other hand, the predicted $K^-K^0$ mass spectra in
Fig.~\ref{fig:D+KK} of the $D^{(*)+}K^-K^0$ modes show peaks close
to threshold, which is due to the near-threshold behavior of the
$F^{KK}_1$ form factor (see Fig.~\ref{fig:KweakFF}). There is no
other clear structure, other than the $B\to D^*$ form factor
effect at larger $q^2$. Because of lower $D^{(*)+}$ reconstruction
efficiencies, the spectra has yet to be measured
experimentally~\cite{Drutskoy:2002ib}, but our predicted spectrum
can be checked soon with more data.

%
%
\begin{figure}
\centerline{\resizebox{0.4\textwidth}{!}{%
\includegraphics{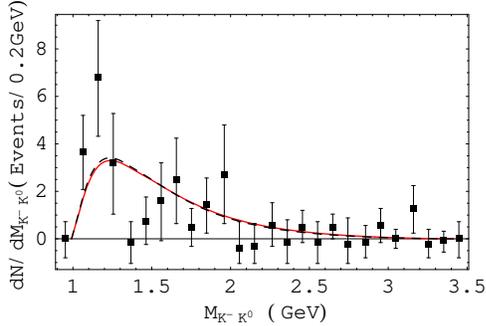}}}
\caption{$B^-\to D^{0}K^-K^0$ spectrum, where solid (dashed) line
is for the MS~(BSW) model, and the data is from
Ref.~\cite{Drutskoy:2002ib}.}\label{fig:d0KKdata}
\end{figure}
%
\begin{figure}
\centerline{\resizebox{0.41\textwidth}{!}{%
\includegraphics{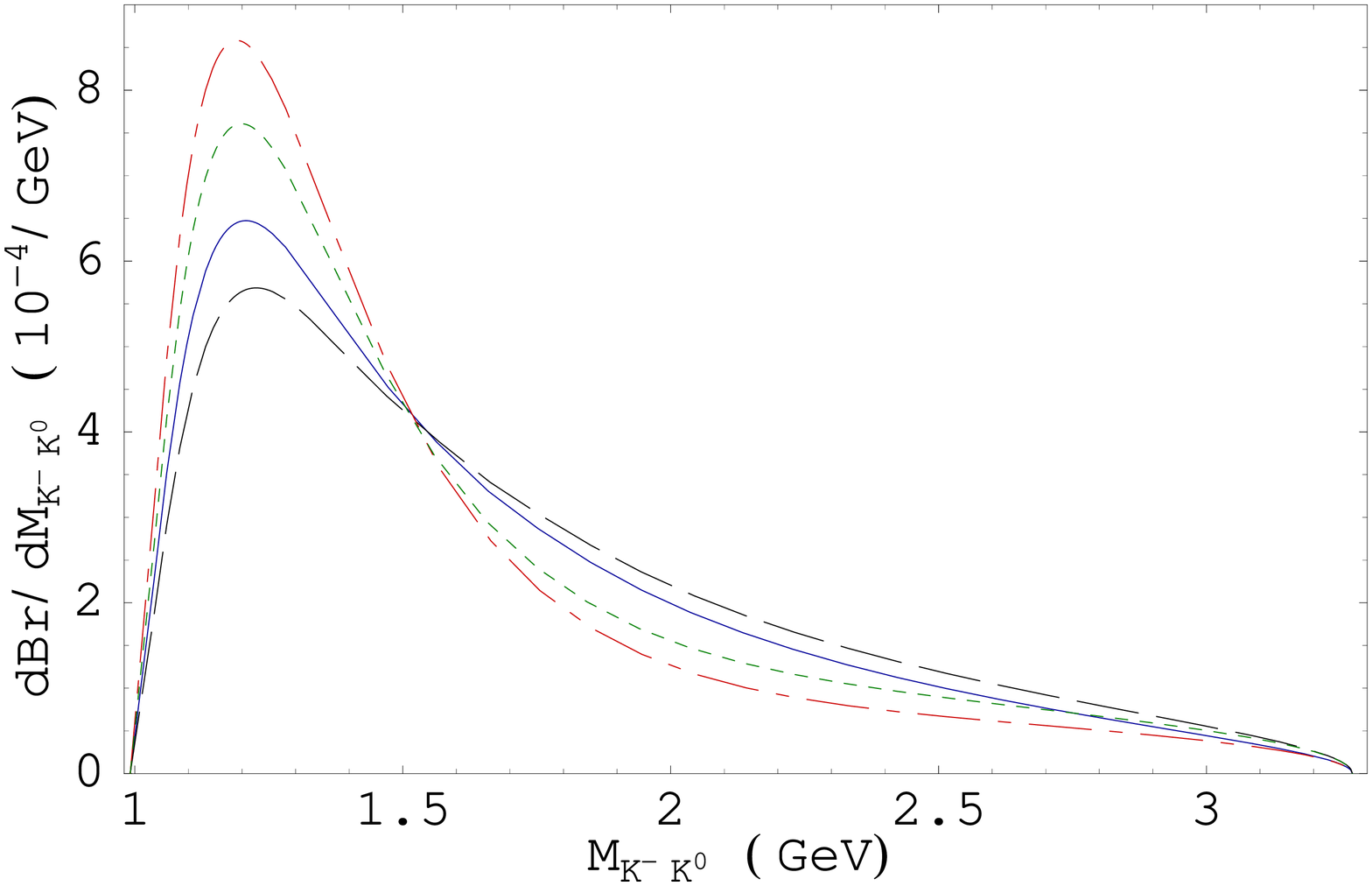}}}
\caption{\small {The $K^-K^0$ mass spectrum for $B^-\to
D^{*0}K^-K^0$, where solid, dot-dashed, dashed and dotted lines
are for MS model with $c_h=11.3,\ -16.1$~GeV$^3$ and BSW model
with $c_h=13.1,\ -18.5$~GeV$^3$,
respectively.}}\label{fig:charmbr2}
\end{figure}

For $B^-\to D^{(*)0} K^-K^0$ which involve the unknown parameters
$c_{w_-}$, $c_h$, we fit to the decay rates to obtain their
values, as shown in Table~\ref{tab:BRD0KK}. The $K^-K^0$ mass
spectra can then be obtained, in particular for the $B^-\to
D^0K^-K^0$ mode, where one can see from Fig.~\ref{fig:d0KKdata}
that the $K^-K^0$ mass spectrum can be roughly described by the
naive form factor model of Eq.~\ref{eq:transitionFF}. The mass
spectrum of $B^-\to D^{*0}K^-K^0$ in Fig.~\ref{fig:charmbr2} can
be checked in experiment.
\begin{table*}
\caption{Fitted values of transition form factor parameters
$c_{w_-}$ and $c_h$, by using the central values of
$D^{(*)0}K^-K^0$ rates. See \cite{Chua:2002pi} for detailed
discussions.}
\label{tab:BRD0KK}       
\begin{center}
\resizebox{0.75\textwidth}{!}{
\begin{tabular}{lccc}
\hline\noalign{\smallskip}
 & $c_{w_-}^{\rm MS(BSW)}$~(GeV$^3$) &  $c_h^{\rm
 MS(BSW)}$~(GeV$^3$)
 & ${\mathcal B}(10^{-4})$~\cite{Drutskoy:2002ib}  \\
\noalign{\smallskip}\hline\noalign{\smallskip}
$B^-\to D^0 K^-K^0$  & $-35.4~(-33.0)$ & --- & $5.5\pm1.4\pm0.8$ \\
$B^-\to D^{*0}K^-K^0$ & $-35.4~(-33.6)$ & $11.3~(13.1)$ or
$-16.1~(-18.5)$
& $5.2\pm2.7\pm1.2$ \\
\noalign{\smallskip}\hline
\end{tabular}
}
\end{center}
\end{table*}

\section{Discussion and Conclusion}
\label{sec:dis-con}

A factorization approach has been used to study three-body
$\overline B\to D^{(*)}K^-K^0$ decays. There are two mechanisms of
kaon pair production, namely current-produced and transition. For
$\overline B^0\to D^{(*)+}K^-K^0$ which involve the former, one
can make use of kaon EM data through isospin rotations. The result
is in good agreement with experiment, which supports
factorization.

The $B^-\to D^{(*)0}K^-K^0$ decays also receive the transition
contribution. The form of these transition form factors are
determined through QCD counting rules, and we fix the parameters
by using the measured $D^0 K^-K^0$ and $D^{*0}K^-K^0$ decay rates.
The predicted mass spectra of the $B^-\to D^0K^-K^0$ mode agrees
well with data and exhibit threshold enhancement as do the
$\overline B {}^0\to D^{(*)+}K^-K^0$ cases. Despite the success in
describing the mass spectrum of the $D^0K^-K^0$ mode, our
treatment of the $B^-\to K^-K^0$ transition form factors may be
oversimplified. Assuming the asymptotic form required by PQCD may
be too strong an assumption, and might have over-enhanced the
contribution from the near-threshold region. More careful study on
other possibilities, such as using pole models for transition via
resonances, would be helpful in clarifying the underlying dynamics
of the $B^-\to D^{(*)0}K^-K^0$ transitions.

Finally, with the feasibility of extracting $K^-K^{*0}$ axial form
factors emboldened by the success in $\overline B\to
D^{(*)}K^-K^0$ decays, $B$ decay data plus factorization have
opened up a new avenue to the study of meson form factors, which
have traditionally been fundamental quantities to many fields in
both nuclear and elementary particle physics. The success of
factorization in $\overline B\to D^{(*)} K^-K^0$ decays urges a
serious study of the underlying mechanism.

\end{document}